\documentclass[aps,prd,twocolumn,superscriptaddress,nofootinbib]{revtex4}

\usepackage{epsfig}
\usepackage{amssymb}
\usepackage{amsmath}
\usepackage{amsfonts}
\usepackage{hyperref}

\newcommand{\be}{\begin{equation}}
\newcommand{\ee}{\end{equation}}
\newcommand{\bea}{\begin{eqnarray}}
\newcommand{\eea}{\end{eqnarray}}
\newcommand{\bi}{\begin{itemize}}
\newcommand{\ei}{\end{itemize}}
\newcommand{\ben}{\begin{enumerate}}
\newcommand{\een}{\end{enumerate}}

\newcommand{\lp}{\left(}
\newcommand{\rp}{\right)}

\def\gsim{\mathrel{\rlap{\lower4pt\hbox{\hskip1pt$\sim$}}
    \raise1pt\hbox{$>$}}}         %greater than or approx. symbol
\def\lsim{\mathrel{\rlap{\lower4pt\hbox{\hskip1pt$\sim$}}
    \raise1pt\hbox{$<$}}}         %less than or approx. symbol

\def \as {\alpha_s}

\begin{document}
\makebox[6.5in][r]{\hfill ANL-HEP-PR-12-100}\\ 
\title{A complete next-to-leading order  QCD description of resonant
$Z'$ production \\ and decay into $t\bar t$ final states}

\author{Fabrizio Caola}
\email{caola@pha.jhu.edu}
\affiliation{Department of Physics and Astronomy, Johns Hopkins 
University, Baltimore, Maryland, USA}
\author{Kirill Melnikov}
\email{melnikov@pha.jhu.edu}
\affiliation{Department of Physics and Astronomy, Johns Hopkins 
University, Baltimore, Maryland, USA}
\author{Markus Schulze}
\email{markus.schulze@anl.gov}
\affiliation{Argonne National Laboratory (ANL), Lemont, Illinois 60439, USA}

\begin{abstract}
We discuss  QCD  radiative corrections to the production 
of a heavy neutral resonance $Z'$ at  the LHC assuming that it decays into 
a $ t\bar t$ final state. 
Compared to previous studies, our computation includes 
top quark decays as well as interference between the $Z'$ signal 
process and the QCD $t \bar t$ background. The interference contribution 
appears for the first  time at next-to-leading 
order QCD and 
requires new one-loop  amplitudes that
are not present when 
signal and background are treated separately.
 We describe  some examples of how QCD radiative corrections may influence 
both the 
exclusion limits and studies of properties of the new resonance, 
once it is discovered. 
\end{abstract}

\maketitle

\section{Introduction}

Searches for physics beyond the Standard Model (BSM) at the LHC are starting to 
reach into the TeV mass range and exclude large fractions of parameter spaces 
of most popular models of new physics.  Many of these exclusion limits--that 
often involve rather complicated final states and extreme 
energies--are based on leading order calculations which are not necessarily adequate in such 
situations.  While, in general, it is unrealistic to expect that theoretical 
description of collider processes that are relevant for  all  
interesting BSM scenarios will be advanced to next-to-leading orded (NLO)
 QCD accuracy, 
this can be  done  for such  extensions of the SM that are considered 
to be sufficiently robust. In fact, such calculations have been reported 
recently for a variety of new physics models in  
Refs.~\cite{Sullivan:2002jt,Gao:2010bb,
Chivukula:2011ng,GoncalvesNetto:2012yt,Zhu:2012um}.

One such robust scenario is an extension of the Standard Model 
by an additional $U(1)$ gauge group which generically leads to the  appearance 
of a massive neutral flavor-conserving spin-one particle which we will 
refer to as $Z'$. 
Physics of such particles has been studied in great detail, and it is beyond the scope of
this paper to discuss all possible model-building aspects. The interested reader can find 
studies in this direction 
in~\cite{Langacker:2008yv,Leike:1998wr,Hewett:1988xc,Carena:2004xs,Appelquist:2002mw}  
and references therein. For recent discussions in the context of the LHC see e.g. Refs.~\cite{Harris:2011ez,Basso:2012ux,deBlas:2012qp,Armillis:2008vp}.

This new $Z'$ particle will, generically, couple to all quarks, 
including the top quark. For this reason, it can be searched for 
in the process $pp \to Z' \to t \bar t$ at the LHC.
In fact, searches for $Z'$ resonances in this  process are well 
under way~\cite{Aad:2012qa,Aad:2012txa,Aad:2012wm,Chatrchyan:2012rq,Chatrchyan:2012cx,Chatrchyan:2012ku}. 
Currently, such searches 
 exclude  $Z'$ with masses smaller than $1.5-2~{\rm TeV}$ where 
the exact  value of  the exclusion  bound depends 
on the assumed couplings of the 
$Z'$ to light quarks and the top quark. 

In this paper we study the QCD corrections  to
the production of a $Z'$ resonance with its subsequent decay to a $t \bar t$ 
pair, without the simplifying assumptions of
previous studies. In particular, we include 
NLO QCD corrections 
to decays 
of top quarks into observable final states, account 
for all spin correlations, and consider an 
interference of the signal $q \bar q \to Z' \to t \bar t$ 
with the gluon-mediated QCD background  $q \bar q \to g^* \to t \bar t$ 
which appears at next-to-leading order  for the first time. 

Our computation can be used to perform realistic studies of 
both the discovery potential of $Z' \to t \bar t$ at the LHC 
and precision studies of the properties of the $Z'$ 
resonance once it is discovered. Hence our computation complements
similar precision studies of $Z'$ collider phenomenology in Drell-Yan final states~\cite{Petriello:2008zr,Coriano:2008wf}.
The potential applications 
of our calculation are illustrated by 
computing $K$-factors for experimental cuts that 
are similar to the ones employed 
by ATLAS and CMS~\cite{Aad:2012qa,Aad:2012txa,Aad:2012wm,Chatrchyan:2012rq,Chatrchyan:2012cx,Chatrchyan:2012ku}
 in their searches for $Z' \to t \bar t$ 
and by studying the kinematic distribution of a suitably defined 
relative azimuthal angle of leptons from semileptonic 
decays of top quarks~\cite{Baumgart:2011wk}   
that can  be used 
to discriminate between vector and axial couplings  of  
$Z'$ to top quarks. 

The rest of the paper is organized as follows. In 
Sec.~\ref{setup} we describe the $Z'$ model 
that we use throughout the paper and comment 
on the general setup of the computation.  In 
Sec.~\ref{qcdnlo} we discuss aspects 
of the NLO QCD calculation that are relevant for our analysis. 
In Sec.~\ref{numerics} we present the numerical results. 
We conclude in Sec.~\ref{conc}.

\begin{table*}
\centering
\begin{tabular}{|c|c|c|c|c||c|c|c|c|}
\hline
& \multicolumn{4}{|c||}{Narrow $Z'$, $\Gamma^{(\mathrm{LO})}_{Z'}=15~$GeV} & 
\multicolumn{4}{|c|}{Broad $Z'$, $\Gamma^{(\mathrm{LO})}_{Z'}=150~$GeV}\\
\cline{1-9}
$(f_1,f_2)$ & (1,1) & (-1,-1) & (1,0) & (0,-1) & (1,1) & (-1,-1) & (1,0) & (0,-1) \\
\hline
\hline 
LO, $\mu=m_{Z'}/2$          &  3.94 & 3.76 & 5.73 & 2.13 &   39.0 & 36.4 & 56.6 & 20.9 \\
LO, $\mu=m_{Z'}$            &  3.70 & 3.53 & 5.39 & 2.00 &   36.8 & 34.3 & 53.4 & 19.6 \\
LO, $\mu=2 m_{Z'}$          &  3.49 & 3.32 & 5.08 & 1.88 &   34.7 & 32.4 & 50.5 & 18.5 \\
\hline
NLO, $\mu=m_{Z'}/2$         &  4.20 & 3.90 & 6.11 & 2.21 &   42.0 & 38.2 & 61.0 & 21.9 \\
NLO, $\mu=m_{Z'}$           &  4.17 & 3.89 & 6.06 & 2.20 &   41.7 & 38.1 & 60.5 & 21.8 \\
NLO, $\mu=2 m_{Z'}$         &  4.12 & 3.86 & 5.98 & 2.18 &   41.3 & 37.8 & 59.9 & 21.7 \\
\hline
$K-$factor, $\mu=m_{Z'}/2$   &  1.07 & 1.04 & 1.07 & 1.04 &    1.08 &  1.05 &  1.08 & 1.05 \\
$K-$factor, $\mu=m_{Z'}$     &  1.13 & 1.10 & 1.12 & 1.10 &    1.13 &  1.11 &  1.13 & 1.11 \\
$K-$factor, $\mu=2m_{Z'}$    &  1.18 & 1.16 & 1.17 & 1.16 &    1.19 &  1.17 &  1.18 & 1.17 \\
\hline
\end{tabular}
\caption{  Cross sections (in fb) 
for $pp\to Z' \to t\bar t\to l \bar l \nu \bar \nu j j + X$ 
at leading and next-to-leading order in perturbative QCD 
for various $Z'$ models at the $14~{\rm TeV}$ LHC.  Results are shown 
for  different choices of the renormalization and factorization scales. We use 
MSTW2008 parton distribution functions \cite{Martin:2009iq}. 
Kinematic cuts on leptons, jets and missing 
energy are specified in Sec.~\ref{numerics}. 
The first column corresponds to 
purely vector $Z'$, the second one to purely axial $Z'$, and 
the third one is the
reference point used in experimental analysis by the ATLAS and CMS collaborations~\cite{Aad:2012wm,Chatrchyan:2012rq}.}\label{resscale}
\end{table*}

\section{Setup}
\label{setup}

In this section we describe the setup of the calculation. 
Our calculation can accommodate any possible renormalizable interaction of the $Z'$ to fermions.
Furthermore, we take both the $Z'$ mass and its width as input parameters of the theory, which
allows us to describe with NLO accuracy a generic $Z'$ model. To present our results however,
we restrict ourselves to the leptophobic top-color model of Refs.~\cite{Harris:1999ya,Harris:2011ez},  
which is the reference model used by the LHC experiments to present their exclusion limits. 

In the context of these leptophobic top-color models, some of parameters 
are correlated. Indeed, the  $Z'$ coupling to  SM particles 
is  described
by the  Lagrangian
\be
\mathcal L  = 
\frac{1}{2} g_1 \cot\theta_H \; Z'^{,\mu} \; 
\left ( J^{L}_\mu + J^{R}_{\mu} \right ),
\label{eq1}
\ee
where $g_1$ is the weak coupling constant and  $\cot \theta_H$ is 
the  mixing parameter that encodes  the overall deviation of 
the $Z'$ couplings to quarks from the electroweak  ones. The left- and 
right-handed currents are defined as 
$J^{\mu}_L =  \bar t_L \gamma_\mu t_L + \bar b_L \gamma_\mu  b_L 
- \bar u_L \gamma_\mu u_L -\bar d_L \gamma_\mu d_L$, 
and 
$J^\mu_R =   
f_1 \left( \bar t_R \gamma_\mu t_R - \bar u_R \gamma_\mu u_R
\right ) +  f_2 \left ( \bar b_R \gamma_\mu b_R  - \bar d_R \gamma_\mu d_R
\right )$.  In the top-color model, one would require 
$f_1 \ge  0$ and/or  $f_2 \le  0$ in the right-handed current for
top-color tilting~\cite{Harris:1999ya}.

Exclusion limits are often presented assuming $f_1 =1$ and $f_2 = 0$ 
which corresponds to the  vector coupling of $Z'$ to top quarks but, 
of course, other choices of these parameters lead to richer physics.  
For this 
reason, we consider four different coupling choices in what follows,
$(f_1,f_2)=(1,1),~(-1,-1),~(1,0),~(0,-1)$.  The first and the second 
cases  correspond to pure vector or pure axial  $Z'$ couplings; the 
remaining choices  describe mixed cases. 

We can use  the interaction Lagrangian Eq.~(\ref{eq1}) to compute the 
total decay width of the $Z'$ boson. At leading order  we obtain 
\be
\begin{split}
\label{LOwidth}
& \Gamma^{\rm LO}_{Z'} = 
\frac{\alpha \cot^2\theta_H m_{Z'}}{8 \cos^2 \theta_W} \Big [
(3+f_1^2+2 f_2^2)
 \\
&    +
\beta_t
\left (  (1+f_1^2)
 -(1-f_1^2-6 f_1) \frac{m_t^2}{m_{Z'}^2} \right ) 
\Big ],
\end{split} 
\ee
where $\beta_t = \sqrt{ 1-4 m^2_t/m^2_{Z'}}$. We therefore 
take   $f_1$, $f_2$, 
$\Gamma^{\rm LO}_{Z'}$ and the mass of the $Z'$ resonance
as free parameters of the theory and derive  the overall coupling 
strength of $Z'$ to quarks from them.  We note that this 
leads to a strong correlation between the strength of couplings to quarks 
and the width of the resonance which may be less pronounced 
in other models.  We will return to this issue later in the 
paper when discussing the importance of the interference between 
signal  and background processes that occurs at NLO QCD.

\begin{table*}
\centering
\begin{tabular}{|c|c|c|c|c||c|c|c|c|}
\hline
& \multicolumn{4}{|c||}{Narrow $Z'$, $\Gamma^{(\mathrm{LO})}_{Z'}=15~$GeV} & \multicolumn{4}{|c|}{Broad $Z'$, $\Gamma^{(\mathrm{LO})}_{Z'}=150~$GeV}\\
\cline{1-9}
$(f_1,f_2)$ & (1,1) & (-1,-1) & (1,0) & (0,-1) & (1,1) & (-1,-1) & (1,0) & (0,-1) \\
\hline
\hline
LO                        &  3.70   &  3.53 & 5.39 &  2.00   &  36.8    & 34.3 & 53.4 & 19.6  \\
\hline
NLO production, $q\bar q$ &  4.93   &  4.74 & 7.17 &  2.67   & 49.4     & 46.3 & 71.6 & 26.4  \\
NLO production, $q g$     & -0.15   & -0.14 &-0.22 & -0.08   & -1.4     & -1.2 & -2.1 & -0.7  \\
NLO production, $g\bar q$ & -0.02   & -0.02 &-0.03 & -0.01   & -0.2     & -0.2 & -0.3 & -0.1  \\
NLO decay                 & -0.60   & -0.57 &-0.87 & -0.31   & -6.1     & -5.6 & -8.8 & -3.2  \\
Interference with QCD     & $\sim0$ & -0.13 & 0.01 & -0.06   &  $\sim 0$ & -1.3 &  0.1 & -0.6  \\
\hline
Full NLO                  &4.17     &  3.89 & 6.06 & 2.20    &  41.7    & 38.1 & 60.5 &  21.8 \\
$K-$factor                &1.13     &  1.10 & 1.12 & 1.10    &   1.13    &  1.11 &  1.13 &  1.11  \\
\hline
\end{tabular}
\caption{Cross sections (in fb) for $pp\to Z' \to t\bar t\to l \bar l \nu \bar \nu j j + X$ 
at the $14~{\rm TeV}$ LHC split
into different contributions, for the reference scale $\mu=m_{Z'}=1.5$~TeV.
We use 
MSTW2008 parton distribution functions. Kinematic cuts on leptons, jets and missing 
energy are specified in Sec.~\ref{numerics}.  
The leading order  process, computed with NLO parameters 
(two-loop $\as$, NLO PDFs and widths) is included in 
the ``NLO production, $q\bar q$'' category.}\label{results}
\end{table*}

As we pointed out already, one of the goals of this paper 
is to present a realistic description of the hadronic $Z' \to t \bar t $ 
production process that combines  QCD radiative corrections 
and top quark decays.  To this end, we have implemented all the 
relevant top quark decay channels. For simplicity, we
will only discuss results for fully leptonic channels 
in what follows.
We use the  narrow width 
approximation for top quark pair production  as described 
in Ref.~\cite{Melnikov:2009dn}. The parametric accuracy 
of this  approximation  is   $\mathcal O(\Gamma_t/m_t)$ 
and its practical reliability is superb as long as the on-shell 
intermediate $t \bar t$ state is kinematically 
allowed~\cite{AlcarazMaestre:2012vp}.

For a complete  NLO QCD analysis of the 
$pp\to Z' \to t\bar t\to XX$ process in the narrow width 
approximation for $t$ and $\bar t$, 
we need the following ingredients:
\begin{itemize}

\item NLO QCD corrections to $pp\to Z' \to t\bar t$ 
for polarized $t$ and $\bar t$. 

\item NLO QCD corrections to the $Z'$ decay width.  Those 
corrections are potentially important to properly describe  shapes of the 
$t \bar t$ invariant mass distribution when ${\cal O}(\alpha_s)$ 
corrections are taken into account. These corrections can be taken 
from the computation of ${\cal O}(\alpha_s)$ corrections 
to $Z\to b\bar b$ decay for massive $b$ quarks reported 
in Ref.~\cite{Kniehl:1989qu}.

\item NLO QCD corrections to top quark decays, 
including NLO QCD corrections to the top quark width 
and the $W$ width. All these
corrections were already implemented 
in the calculation of $pp \to t \bar t$ 
reported in Ref.~\cite{Melnikov:2009dn} and we take them from there.
 
\item NLO QCD corrections to the 
QCD background process $pp\to t\bar t$. We use the implementation described 
in Ref.~\cite{Melnikov:2009dn}. 

\item Interference  of the background $q \bar q  \to t \bar t$ 
and signal  $q \bar q \to Z \to t \bar t$ processes that appears at
NLO QCD for the first time.

\end{itemize}

\begin{figure*}
\centering
\includegraphics[scale=0.7]{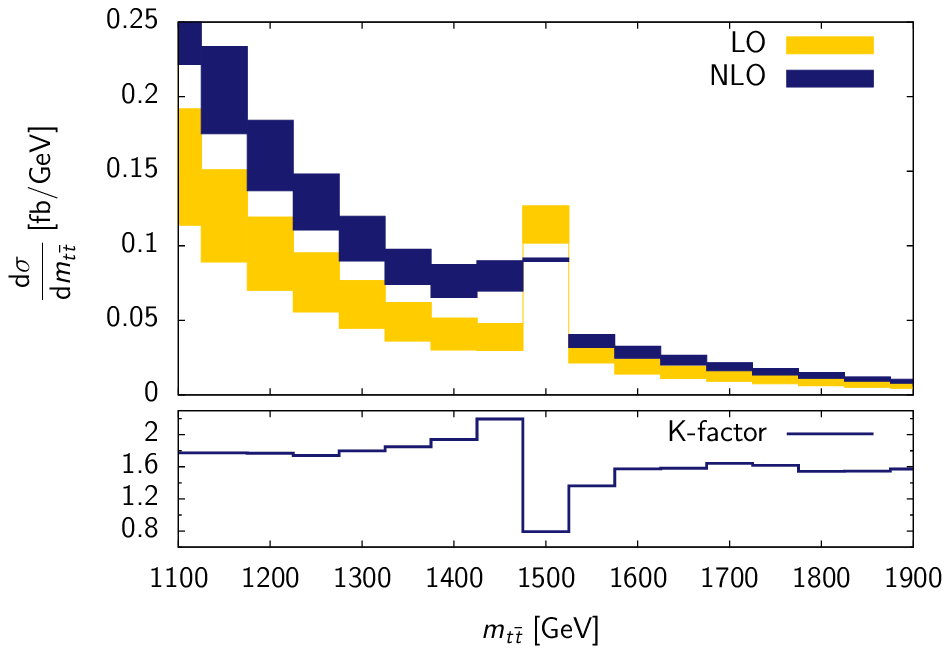}
\includegraphics[scale=0.7]{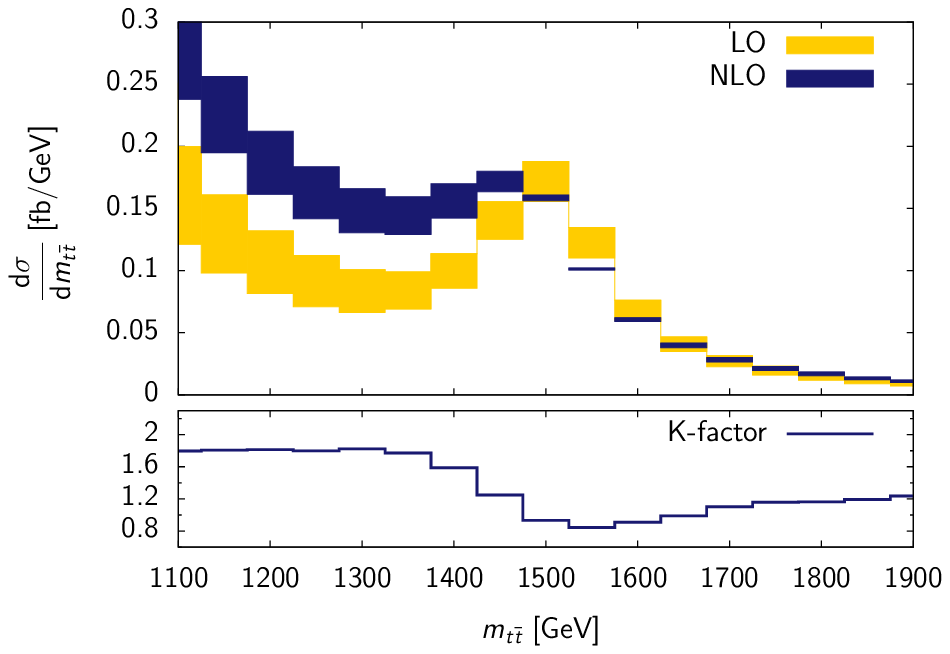}
\caption{Top-antitop invariant mass distribution in the reaction $pp \to t \bar t$ 
for $m_Z' = 1.5~{\rm TeV}$ at the $14~{\rm TeV}$ LHC at leading and next-to-leading 
order in perturbative QCD.   
 Left panel: narrow $Z'$, $\Gamma_Z' = 15~{\rm GeV}$ and  $f_1 = 1, f_2= 0$.  Right panel: 
broad $Z'$, $\Gamma_Z' = 150~{\rm GeV}$ and $f_1 = -1, f_2= -1$.
We use MSTW2008 parton distribution functions.
Kinematic cuts on leptons, jets and missing 
energy are specified in Sec.~\ref{numerics}. 
The $K$-factors are computed
for the central scale choice $\mu=m_{Z'}$.
}\label{mtt_narrow}
\end{figure*}

\section{NLO QCD corrections to $pp\to Z' \to t\bar t$}
\label{qcdnlo}

In this section, we describe some details pertinent to the 
computation of NLO QCD corrections to $pp \to Z' \to t \bar t$. 
Through $\mathcal O (\as)$, we can 
express  the $q\bar q\to Z' \to t\bar t$ amplitude 
as
\be
\label{oneloopamplzp}
\begin{split}
& \mathcal A_{Z'} = 
\delta_{i_q i_{\bar q}}\delta_{i_t i_{\bar t}} A_{Z'}^{(0)} 
+ g_s^2 \Big [ 
\frac{N_c^2-1}{2N_c}  \delta_{i_{q} i_{\bar q}}
\delta_{i_t i_{\bar t}}\mathcal C_1 
\\
& 
+ \lp \delta_{i_q i_{\bar t}}\delta_{i_t i_{\bar q}}
-\frac{1}{N_c} \delta_{i_{q} i_{\bar q}}\delta_{i_{t} i_{\bar t}}\rp \mathcal C_2
\Big ],
\end{split} 
\ee
where $N_c = 3$ is the number of colors. 
The one-loop contribution  to the amplitude ${\cal A}_{Z'}$ is  given by the two 
color-stripped primitive amplitudes ${\cal C}_1$ and ${\cal C}_2$.  
$\mathcal C_1$ describes the amplitude 
with the same color flow as the tree-level one
$A_{Z'}^{(0)}$; it is computed from vertex corrections 
to the production and decay stage of the process 
$q \bar q \to Z' \to t \bar t$. It also receives contributions 
from the required renormalization constants. 
The function  $\mathcal C_2$ 
receives contributions  from box diagrams where a gluon 
and a $Z'$ connect the initial $q \bar q$ and the final $t \bar t$ states.

It is clear from Eq.(\ref{oneloopamplzp}) that 
 the amplitude $\mathcal C_2$ corresponds to the 
color-octet flow  in the $s$-channel 
while the amplitude $\mathcal C_1$  and the tree-level 
amplitude $A_{Z'}^{(0)}$ correspond to the color-singlet flow 
in the $s$-channel.   For this reason, $\mathcal C_2$ does not 
contribute to the production cross section of $Z'$ if we restrict 
ourselves to the underlying process $q \bar q \to Z' \to t \bar t$. 
On the other hand, we note that the color-octet amplitude 
can interfere with the background process 
$q \bar q \to g^* \to  t \bar t$. Such interference is not 
expected to be large for the narrow resonance but it is not clear 
\emph{a priory} if it is negligible for broader resonances. 

To be specific, consider the pure QCD scattering amplitude for 
$q \bar q \to t \bar t$ through one loop. Its color decomposition 
reads~\cite{Melnikov:2009dn}
\be
\label{oneloopamplqcd}
\begin{split} 
& \mathcal A_{\rm QCD}= g_s^2 
\lp\delta_{i_q i_{\bar t}}\delta_{i_t i_{\bar q}}
-\frac{1}{N_c} \delta_{i_q i_{\bar q}}\delta_{i_t i_{\bar t}}\rp A_{\rm QCD}^{(0)} 
\\
& + g_s^4 
\left[
\delta_{i_q i_{\bar t}} \delta_{i_t i_{\bar q}} \mathcal B_1 -\frac{1}{N_c} 
\delta_{i_q i_{\bar q}}\delta_{i_t i_{\bar t}} \mathcal B_2
\right],
\end{split} 
\ee
where  tree- and one-loop contributions are separated.  It follows 
from Eqs.~(\ref{oneloopamplzp},\ref{oneloopamplqcd}) that 
the tree-level QCD amplitude $A_{\rm QCD}^{(0)} $ can interfere 
with ${\cal C}_2$.  We can also see from Eq.~(\ref{oneloopamplqcd}) that, at 
one loop,  ${\cal A}_{\rm QCD}$ 
contains the amplitude ${\cal B}_2$ 
whose  color flow is equivalent to the 
color-singlet exchange in the $s$-channel. For this reason, 
${\cal B}_2$  interferes with the tree-level amplitude 
for the $Z'$ production  $A_{Z'}^{(0)} $, 
but ${\cal B}_2$ does not contribute to the pure QCD 
process $q \bar q \to t \bar t$ due to the color structure 
of the $q \bar q \to g^* \to t \bar t$ amplitude at leading order. 
Therefore if  we consider the 
physical process $q \bar q \to t \bar t$  that can be 
mediated by both gluon and $Z'$ exchanges and if we 
allow for the interference effects between the two 
processes--that \emph{a priori} are not obviously negligible for 
broad $Z'$ resonances--two amplitudes beyond what 
has been considered in the literature so far are required. 
We compute those amplitudes and include the effects of the interference 
between signal and background processes in the discussion below.

We note that we neglect contributions of 
the one-loop ``anomaly-like'' diagrams $gg \to Z' \to t \bar t$ that 
are mediated  by quark loops.  Such contributions  are 
expected to be negligible  for several reasons.  First, 
the gluon flux at center-of-mass energies relevant for $Z'$ production 
is expected to be about one-tenth of the $q \bar q$ flux. Second,
the anomaly-free nature of the effective Lagrangian 
Eq.~(\ref{eq1}) ensures that $gg \to Z'$ vanishes identically if 
all quarks are massless. The quality of the massless approximation 
for top quarks in the $gg \to Z'$ amplitude 
is controlled by  the ratio  of the top 
quark mass to the $Z'$ mass which is quite small for 
  values of the $Z'$ mass that are of interest 
to us.  We conclude that the double suppression 
caused by the smallness of the gluon flux and the validity of the massless 
approximation for top quarks allows us to neglect the ``anomaly'' contribution. 

We also neglect the contribution coming from exchange of the
standard model $Z$ and
photon in the $q\bar q$ annihilation channel, as they are expected to be tiny~\cite{andreas} in the
invariant mass region $m_{t\bar t}\approx m_{Z'}\sim 1$~TeV which is discussed in this
paper. We note that if required by some particular study, the inclusion of such interference 
contributions in our computation is straightforward since
we work at the amplitude level and all coupling and masses that we use are free parameters. 

As a last remark, we would like to elaborate on the QCD corrections 
to the $Z'$ width. In general, we write the $Z'$ propagator in the unitary 
gauge as 
\be
{\cal D}^{Z'}_{\mu \nu}  
= \frac{\mathrm{i}}{k^2 - m^2_{Z'} + \mathrm{i} m_{Z'} \Gamma_{Z'}} 
\left ( - g_{\mu \nu} + \frac{k_\mu k_\nu}{m_{Z'}^2} \right ).
\label{eqBW}
\ee
The longitudinal part of the $Z'$ propagator is not important since 
it always couples to the massless quark line on the ``production side'' 
of the process.   When NLO QCD corrections to the production 
of $Z'$ are considered, it is essential to include corrections 
to $\Gamma_Z'$ in the denominator of the $Z'$ propagator of 
Eq.~(\ref{eqBW}). To see this, consider production of a {\it narrow} 
$Z'$, such that $\sigma(pp \to Z' \to t \bar t) \approx 
\sigma(pp \to Z') \times {\rm Br}(Z' \to t \bar t)$.  To correctly 
compute the change in the branching ratio of $Z' \to t \bar t$ 
at NLO QCD, we need to account for  corrections to the 
partial width $Z' \to t \bar t$ that we do by computing QCD corrections 
as described above {\it and} for  the
corrections to the  total width $\Gamma_{Z'}$ that we have to compute 
separately 
and include in a Breit-Wigner propagator.  We 
read off the NLO QCD corrections to the $Z'$ width 
from the calculation  of QCD 
corrections to the decay rate  $Z\to b\bar b$ described 
in Ref.~\cite{Kniehl:1989qu}, where they are given as functions 
of $m_b/m_{Z}$. For numerical examples that we consider below 
$m_t/m_{Z'} \ll  1$, and in that limit corrections 
to the width become simple. Indeed, if we use the massless 
top quark limit,
the NLO width 
is just $(1 + \alpha_s/\pi)$ times the LO width. 
Nevertheless, in our calculation,
we retain  full $m_t/m_{Z'}$ dependence of the width,
 following  Ref.~\cite{Kniehl:1989qu}.

\section{Numerical results}
\label{numerics}

In this section we illustrate our computation by presenting some numerical results. 
We consider the LHC with $\sqrt{s} = 14$~TeV center-of-mass energy.
We consider semileptonic decays of both $t$ and $\bar t$. 
 We use the anti-$k_t$ jet algorithm with
$R=0.5$ and impose  cuts on the final state particles
that are motivated by current $Z'$ searches 
\cite{Aad:2012wm,Chatrchyan:2012rq}. 
In particular, we require $p_{\perp,\rm lep} < 20~\text{GeV}$, 
$|\eta_{\rm lep}| < 2.5$, $p_{\perp, \rm jet} < 30~\text{GeV}$, 
$|\eta_{\rm jet}| < 2.5$.
We choose $m_{Z'}=1.5~$TeV, which is close 
to the current exclusion limits for our choices 
of the $Z'$ couplings. 
We use the MSTW2008 parton distribution functions~\cite{Martin:2009iq}.

As pointed out in the Introduction, 
instead of specifying the mixing angle $\theta_H$ 
we find it more convenient
to use the leading order  
width as an input parameter. We choose  two reference values 
$\Gamma^{\rm LO}_{Z'}=15~{\rm GeV}$ (narrow $Z'$) 
and $\Gamma^{\rm LO}_{Z'}=150~{\rm GeV}$ (broad $Z'$). 
We also choose $m_{Z'}$ as a reference value  for the  
renormalization and factorization scales and show 
results for three choices $\mu = m_{Z'}/2,m_{Z'}$ and $2 m_{Z'}$.   
 We note that, most likely,  this procedure   overestimates
the theoretical uncertainty  
since it implies a scan over a very broad range of 
scales, between $\mu = 750~\text{GeV}$ and $3~\text{TeV}$.

\begin{figure*}
\includegraphics[scale=0.7]{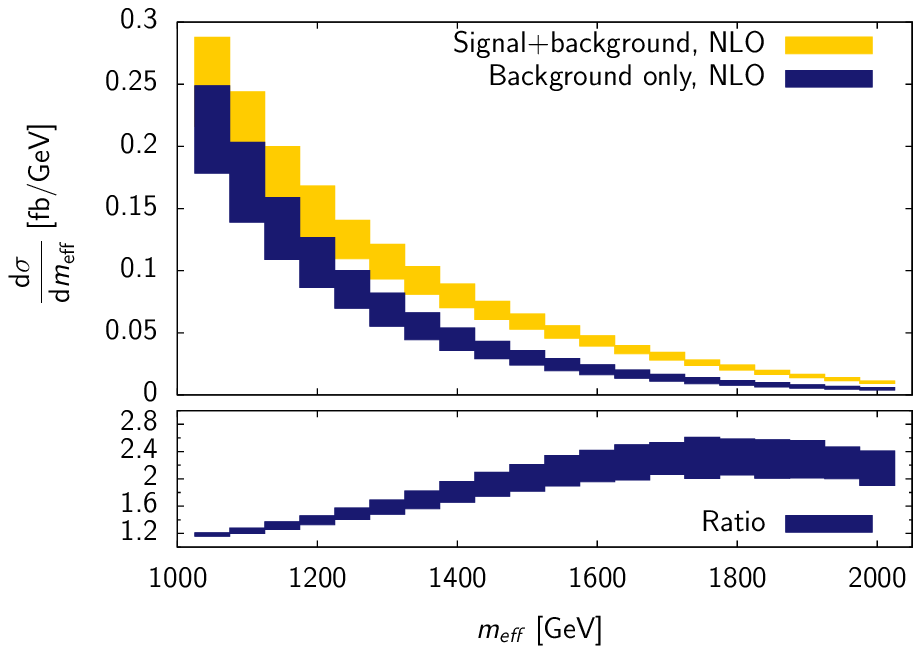}
\includegraphics[scale=0.7]{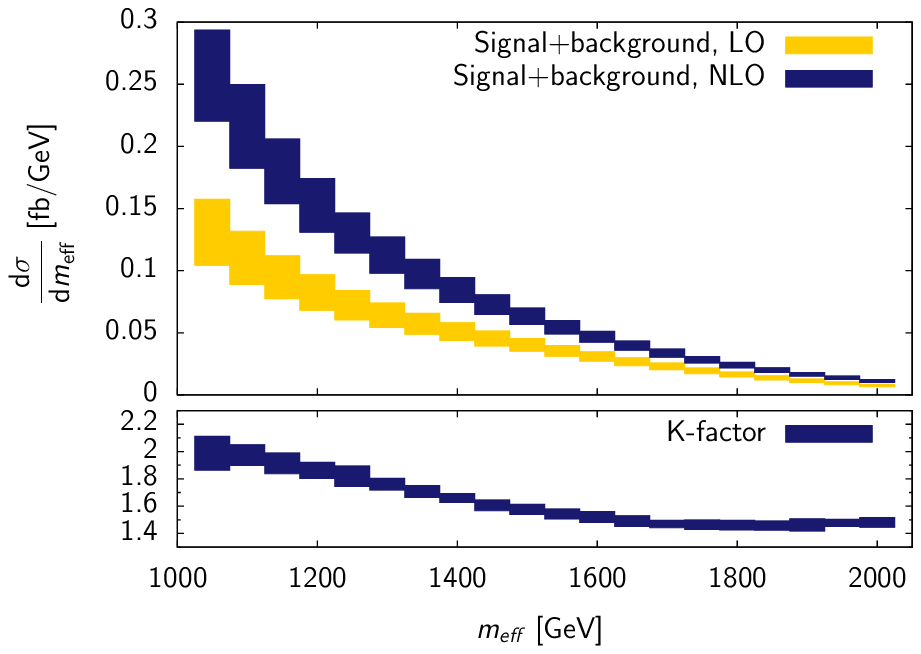}
\caption{ Distribution of the effective mass defined 
as the sum of the transverse missing energy and $H_\perp$
for broad $f_1 = f_2 = -1$  $Z'$ at the $14$~TeV LHC. 
We use MSTW2008 parton distribution functions.
Kinematic cuts on leptons, jets and missing 
energy are specified in Sec.~\ref{numerics}. 
}\label{meff}
\end{figure*}

We summarize our results  for production cross sections 
in the dilepton channel for both   
narrow and broad $Z'$ resonances in Table~\ref{resscale}.  The inclusion 
of NLO QCD corrections leads to extremely stable results 
under changes of the renormalization and factorization scales.
The residual scale dependence is 
of the order of $1\%$. 
The mild  scale dependence is 
in part  due to the fact that the LO result does not explicitly 
depend on $\alpha_s$ so that the scale variation only comes from 
parton distribution functions.   For $\mu = m_{Z'}$--we find that 
the NLO QCD corrections enhance the production cross section by about 
$10\%$, so the corresponding $K$-factor is $1.1$. This 
value is somewhat lower than $K \approx 1.3$, which was reported in 
Ref.~\cite{Gao:2010bb}, and we attribute this difference to our use 
of MSTW2008 parton distribution functions, kinematic cuts on top quark 
decay products and inclusion of radiative corrections to top quark decays.   
We note that we can reproduce the results for the NLO QCD $K$-factors 
reported in Ref.~\cite{Gao:2010bb} if we use the setup for the calculation 
described in that reference. 

To understand the importance of various NLO QCD effects, 
we display in Table~\ref{results} 
all the different contributions to the NLO cross section for $\mu=m_{Z'}$. 
The ``production'' contribution corresponds 
to NLO QCD 
corrections to  various  production processes 
$ pp \to Z'\to t \bar t$ and leading 
order  top quark decays. 
We include the leading order  process $q \bar q \to Z' \to t \bar t$  
computed with the QCD-corrected values 
of $\Gamma_{Z'}$ and $\Gamma_t$  in  the production contribution. 
 The ``decay'' contribution corresponds to leading order 
$q \bar q \to Z' \to t \bar t$ and NLO QCD decays of $t$ and $\bar t$. 
The negative correction from the decay somewhat  compensates the positive change in the leading order 
processes caused by using $\Gamma_t^{\rm NLO}$ there, 
but the cancellation 
is not complete because of the fiducial volume cuts.   
Finally, the last contribution displayed in Table~\ref{results} 
refers to the interference between  QCD  amplitudes and  
$Z'$ amplitudes that arises at NLO QCD as  we already discussed.

An interesting feature of these  
results is that although the $q \bar q $ channel gives the largest contribution
and it has an associated $K$-factor of about $1.3$, similar 
to the result adopted in a recent 
analysis by the ATLAS and CMS 
collaborations \cite{Aad:2012wm,Chatrchyan:2012rq}, 
all other  contributions to the $K$-factor are 
negative.\footnote{We mention that ATLAS and CMS use these $K$-factors for $7$ and $8$~TeV LHC, whereas 
we have only shown results for $14$~TeV LHC. We have, however, calculated 
the $K$-factor for $7$~TeV LHC for the narrow $Z'$ with a mass of $1.5$~TeV  
for the coupling choice $f_1 = 1, f_2 = 0$ and fiducial cuts 
as described in the text; we found $K = 1.1$ for $\mu = m_{Z'}$.} 
The net 
effect of these corrections is the reduction of the $K$-factor 
to a smaller value, $ K \sim 1.1$. 

We also note that  the relevance of the 
interference between signal and background 
production processes depends on the details of the model. For the 
$Z'$ model that we employ in this paper, the 
interference  can reach a few percent, but it can be 
larger  in general.  The interference is absent for pure vector couplings 
of $Z'$ due to Furry's theorem, and it is maximal for pure axial couplings. 
For the model specified by the Lagrangian Eq.~(\ref{eq1}) 
the magnitude of interference contributions can be as large 
as the production  contribution in the $qg$ channel.

The interference 
effects become more relevant for broader $Z'$ resonances. 
While this is an expected feature of the result, 
it is perhaps not as apparent 
in Table~\ref{results} as it should be, so we comment on how 
to properly interpret what is seen there.  
Indeed, a glance  at Table~\ref{results} suggests that 
the relative importance of the interference is similar 
for narrow and broad $Z'$. However, this is an artifact 
of our model choice where the growth of the $Z'$ width is 
correlated with the growth of the $Z'$ couplings to quarks. 
Since the production cross sections for 
$pp \to Z' \to t \bar t$   is {\it quadratic}  in those couplings, 
while the interference is {\it linear}, the fact that both the cross section 
and the interference increase by an order of magnitude for 
the broad $Z'$ compared to narrow $Z'$ suggests that the relative 
importance of the interference increases by a factor of $3$.  Hence, 
we conclude that such interference 
effects can be important for broad resonances, especially if the width
of the resonance is only weakly correlated with its coupling to 
quarks. We note that this interference contribution can play a
role in such observables as the top quark 
forward-backward asymmetry at the Tevatron.  Although a flavor-conserving
$Z'$ resonance in the $s$-channel is already ruled out as a possible
explanation for the observed asymmetry, it is interesting to note
that the NLO QCD interference contribution to the asymmetry in our reference
model can be as large  as 20\% of the leading order  asymmetry generated 
by $Z'$ exchange.  The size of the effect suggests that  it is 
worthwhile  to study it for more 
realistic models that explain  the Tevatron asymmetry. 
We leave this for future work.

\begin{figure*}
\includegraphics[scale=0.7]{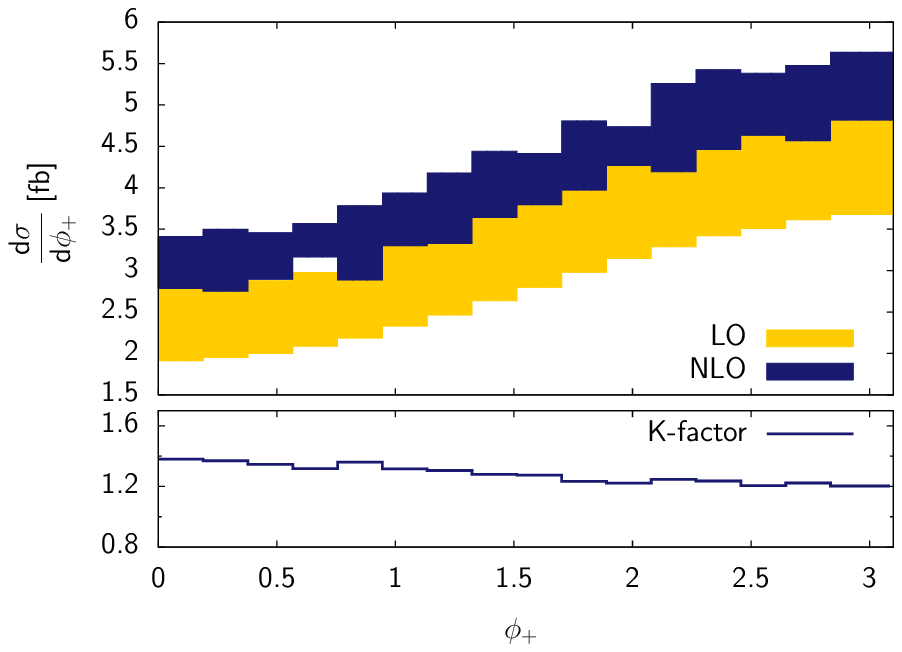}
\includegraphics[scale=0.7]{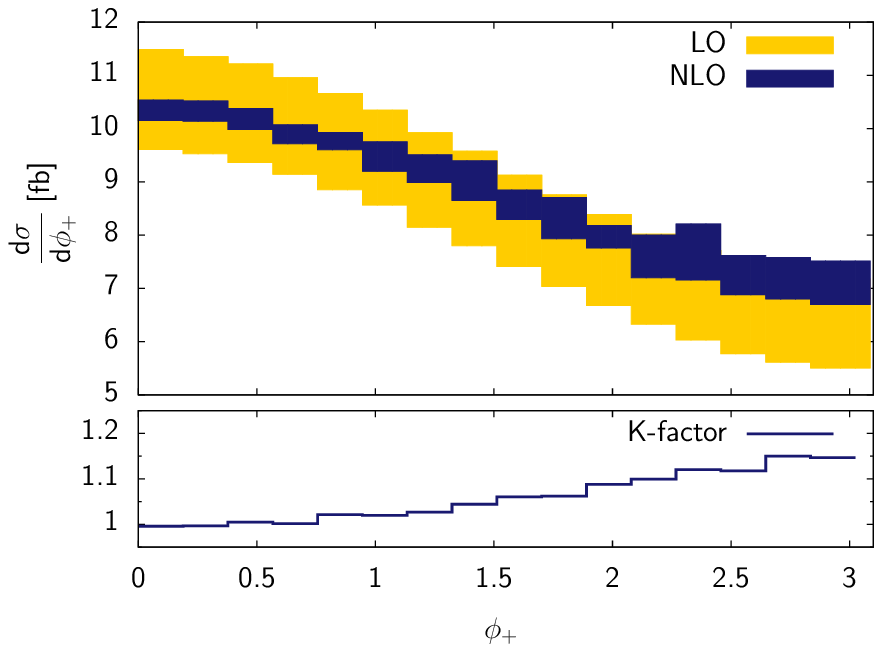}
\caption{Distributions of   $\phi_+$ variable, introduced in~\cite{Baumgart:2011wk}, 
at leading and next-to-leading order in  QCD at $14$ TeV LHC.
Both signal and background are included in the $|m_{t\bar t}-m_{Z'}|<100$~GeV window. 
Left panel: narrow (1,0) $Z'$ model. Right panel: broad (-1,-1) $Z'$.
We use MSTW2008 parton distribution functions.
Kinematic cuts on leptons, jets and missing 
energy are specified in Sec.~\ref{numerics}. 
}\label{mattangle}
\end{figure*}

We now proceed to the discussion of selected differential distributions.
We begin with the 
 distribution in the $t\bar t$ invariant mass. While the $t \bar t$ 
invariant mass is difficult to reconstruct in the dilepton channel, 
it is still instructive to look at the generic features of this distribution.
 Such distributions for narrow and broad $Z'$ are shown 
in Fig.~\ref{mtt_narrow} where we also include the contribution 
from the $pp \to t \bar t$ background process at leading and next-to-leading 
order in perturbative QCD. We employ the reference parameter choices  
$(f_1,f_2)=(1,0)$ and $(f_1,f_2)=(-1,-1)$, respectively for the 
narrow and broad $Z'$ cases reported in Fig.~\ref{mtt_narrow}. The 
$m_{t \bar t}$ invariant mass distribution shows expected features--the 
$Z'$ peak, which is very prominent at leading order, becomes 
broader and less prominent at NLO QCD since a significant fraction 
of events is shifted to  
lower $m_{t \bar t}$  values because of the final state radiation.  
As a result, the $K$-factor changes significantly around 
the peak position, for both 
narrow and broad $Z'$.  This rapid change in the $K$-factor may be invisible 
for the narrow $Z'$, given that the resolution of  $m_{t \bar t}$ is close 
to $100$ GeV  but it may become important for broader resonances. 
Indeed, as follows from the right panel of Fig.~\ref{mtt_narrow}, the $K$-factor 
changes by almost a factor of $2$, from $K = 1.8$ at $m_{t \bar t} = 1300~{\rm GeV}$ to 
$K = 0.8$ at $m_{t \bar t} = 1600~{\rm GeV}$ despite the fact that the average 
$K$-factor is relatively modest. 
Note that the 
 $K$-factor at lower invariant mass values--which are more accessible 
experimentally--is 
actually quite significant and it almost doubles the expected 
cross section there. As a further illustration of this point, we note that 
the peak of 
the NLO QCD $m_{t \bar t}$ distribution  for a broad $Z'$ shown in the 
right panel of  Fig.~\ref{mtt_narrow}
is located in  a different 
bin when compared to the leading order $m_{ t \bar t}$ distribution.

Our calculation allows us to compute any kinematic distribution that 
is relevant for $Z'$ production through NLO QCD and, in particular, 
those that are defined using momenta of top quark decay products. 
For illustration purposes, we show two such distributions. 
In Fig.~\ref{meff} the distribution of the effective mass for 
events with and without (broad) $Z'$ is shown. All NLO QCD effects
that are described above are included in the calculation.  
Understanding the effective mass distribution 
is of paramount importance for the experimental analysis,  where it is 
used to extract information about the $Z'$ mass in channels where direct 
reconstruction of $m_{t \bar t}$ is not possible.  It follows from the 
the right panel of Fig.~\ref{meff} that the $K$-factor is a rapidly changing 
function 
of $m_{\rm eff}$. This is a consequence of the fact that 
the $m_{\rm eff}$ distribution receives contributions from signal 
and background processes which  have markedly  different 
dependence on $m_{\rm eff}$ and very different $K$-factors.

As another example, we show a kinematic distribution that can 
be constructed if lepton momenta are accurately 
measured.  As was pointed out in Ref.~\cite{Baumgart:2011wk}, 
such distributions are  useful for exploring   the Lorentz structure of the 
couplings of the new resonance to top quarks.
 Since our NLO QCD computation includes the spin correlations exactly, 
we can extend   the discussion of the relevant kinematic 
distributions by including NLO QCD corrections.

We consider the  kinematic distribution of the 
azimuthal angle $\phi_+$,  which 
is constructed in the following way \cite{Baumgart:2011wk}. 
We work in the center-of-mass frame\footnote{Because of the missing 
energy in dilepton events, it is difficult, but not impossible, 
to reconstruct the $t \bar t$ center-of-mass 
frame. See the discussion of this 
point in Ref.~\cite{Baumgart:2011wk}.}   of $t$ and $\bar t$. 
In that frame the directions of the incoming protons are not back to 
back; we denote 
them as $n_1$ and $n_2$.  We define  a three-vector 
$\vec n_{12} = (\vec n_1  - \vec n_2)/|\vec n_1  - \vec n_2|$
and form a coordinate system by taking the $z$ axis to be 
the direction of the top momentum $\vec n_t$, the $x$ axis to
be $\vec n_{12}$ and the $y$ axis to be $\vec n_t \times \vec n_{12}$.
We now write the momenta of the lepton and 
the antilepton in that reference frame as
$p_{l,\bar l} = E_{l,\bar l} \left ( 1, \sin \vartheta 
\cos \varphi_{l,\bar l}, \sin \vartheta 
\sin \varphi_{l , \bar l}, \cos \theta_{l, \bar l} \right )$ and define the angle $\phi_+$ 
as 
$
\phi_+ = \cos^{-1} \left ( \cos \varphi_{l} \cos \varphi_{\bar l} - \sin \varphi_{l} \sin \varphi_{\bar l} 
\right ).
$
As was pointed out in Ref.~\cite{Baumgart:2011wk}, 
the $\phi_+$ distribution can be used to discriminate 
between vector and axial couplings of $Z'$ to a $t \bar t$ pair.  

We show  the results of the computation of the $\phi_+$  distribution
in  Fig.~\ref{mattangle} where 
two scenarios ({\it i}) narrow $Z'$ with $f_1 = 1$ and $f_2=0$ and ({\it ii})
broad $Z'$ with $f_1= f_2 = -1$ are shown.  
The distributions 
contain $t$ and $\bar t$ 
from both the 
QCD background  process and from the $Z'$ production.  To reduce the contribution 
of the background, we require that the invariant mass of the $t \bar t$ 
system satisfies  the constraint 
$|m_{t\bar t}- m_{Z'}| < 100$~GeV. With this cut, the signal-to-background 
ratio is roughly $0.5$ for the narrow resonance and is close to a factor of 2
 for the broad resonance 
case.  It follows from Fig.~\ref{mattangle} that the shape of this distribution is fairly stable 
against NLO QCD radiative corrections although in case ({\it ii}) (right panel)
the QCD corrections make the distribution more flat. Since 
a small admixture of a vector coupling will have 
a similar effect on that distribution, 
it is essential to accommodate radiative corrections 
for the precise measurement of $Z't \bar t$ couplings.

\section{Conclusion} 
\label{conc}

In this paper, we discussed  NLO 
QCD radiative corrections to the production 
of a 
$Z'$ boson in the reaction $pp \to Z' \to t \bar t$. In contrast to 
previous studies, our computation  includes 
interference effects of  the $Z'$ production  process and 
the QCD background process 
that may become relevant for broad resonances. In addition, we  
include  radiative corrections to  top quark decays, keeping 
all spin correlations intact.  We find that 
the NLO QCD corrections increase leading order cross sections 
by a modest  amount that is 
somewhat smaller 
than what has been adopted in recent studies by  
the ATLAS and CMS collaborations \cite{Aad:2012wm,Chatrchyan:2012rq}. 
We also find that 
radiative corrections to the  $t \bar t$ invariant 
mass distribution and to  the effective mass distribution  
are significant and strongly dependent  on $m_{t \bar t}$ and $m_{\rm eff}$, respectively.  For 
broader resonances, they increase the number of events with a
smaller invariant mass of the 
$t \bar t$ system making heavier resonances more accessible experimentally.

Our calculation is useful for  
understanding the Lorentz structure of the $Z'$ interaction 
with the $t \bar t$ pair, with high accuracy,  once the resonance is 
discovered. We have illustrated this  by computing 
the distribution  in the relative angles 
of leptons and antileptons, defined as 
suggested in Ref.~\cite{Baumgart:2011wk}.  This variable was 
found~\cite{Baumgart:2011wk} to be 
a good discriminator between vector and axial 
couplings of $Z'$ to top quarks; our computation allows one to check if 
this conclusion remains valid when  NLO QCD corrections are included. 
We found  that, in the case of a narrow $Z'$, 
 the effect of QCD radiative corrections is to rescale 
this distribution by an overall factor without changing its shape while the situation with 
broader $Z'$ is somewhat more complicated. At any rate, our calculation makes it possible 
to analyze any $Z'$ scenario with NLO QCD accuracy. We look forward 
to  the confrontation of our computation with LHC data. 

{\bf Acknowledgments} 
This research is partially supported by the US NSF under 
grants PHY-1214000 and by the
US DOE under grants DE-AC02-06CD11357.
Calculations reported in this paper were performed on the Homewood High Performance Cluster 
of Johns Hopkins University.

%\bibliographystyle{unsrt}
%\bibliography{zprime}

\end{document}